# Fastest Distributed Consensus Problem on Fusion of Two Star Networks

Saber Jafarizadeh

Sharif University of Technology

Department of Electrical Engineering

Azadi Ave, Tehran, Iran

jafarizadeh@ee.sharif.edu

saber.jafarizadeh@gmail.com

**Abstract**

Finding optimal weights for the problem of Fastest Distributed Consensus on networks with different topologies has been an active area of research for a number of years. Here in this work we present an analytical solution for the problem of Fastest Distributed Consensus for a network formed from fusion of two different symmetric star networks or in other words a network consists of two different symmetric star networks which share the same central node. The solution procedure consists of stratification of associated connectivity graph of network and Semidefinite Programming (SDP), particularly solving the slackness conditions, where the optimal weights are obtained by inductive comparing of the characteristic polynomials initiated by slackness conditions. Some numerical simulations are carried out to investigate the trade-off between the parameters of two fused star networks, namely the length and number of branches.

**Keywords:** Fastest distributed consensus, Weight optimization, Sensor networks, Second largest eigenvalue modulus, Semidefinite programming, Distributed detection,

## I. INTRODUCTION

Distributed computation in the context of computer science is a well studied field with an extensive body of literature (see, for example, [1] for early work), where some of its applications include distributed agreement, synchronization problems, [2] and load balancing in parallel computers [3, 4].

A problem that has received renewed interest recently is distributed consensus averaging algorithms in sensor networks and one of main research directions is the computation of the optimal weights that yield the fastest convergence rate to the asymptotic solution [5, 6, 7], known as Fastest Distributed Consensus averaging Algorithm, which computes iteratively the global average of distributed data in a sensor network by using only local communications. Moreover algorithms for distributed consensus find applications in, e.g., multi-agent distributed coordination and flocking [8, 9, 10, 11], distributed data fusion in sensor networks [12, 13, 6], fastest mixing Markov chain problem [14], clustering [15, 16] gossip algorithms [17, 18], and distributed estimation and detection for decentralized sensor networks [19, 20, 21, 22, 23].

Most of the methods proposed so far usually avoid the direct computation of optimal weights and deal with the Fastest Distributed Consensus problem by numerical convex optimization methods and in general no closed-form solution for finding Fastest Distributed Consensus has been offered so far except in [1, 24, 25], where for the path network the conjectured optimal weights [3] has been proved in [1], and in [25], the author has solved Fastest Distributed Consensus problem analytically for Path network using semidefinite programming without any assumption or conjecture, also in [24] the author proposes an analytical solution for Fastest Distributed Consensus problem over complete cored and symmetric star networks.

Here in this work, we aim to solve Fastest Distributed Consensus problem for the fusion of two symmetric star networks called Two Fused Star (TFS) network or in other words a network consists of two different symmetric star networks which share the same central node, by means of stratification and Semidefinite Programming (SDP), particularly solving the slackness conditions, where the optimal weights are obtained by inductive comparing of the characteristic polynomials initiated by slackness conditions. The simulation results confirm that the distributed consensus algorithm with optimal weights converges substantially faster than the one with other simple weighting methods, namely maximum degree, Metropolis and best constant weighting methods; moreover we have investigated the tradeoff between the parameters of network and convergence rate by numerical results.

The organization of the paper is as follows. Section II is an overview of the materials used in the development of the paper, including relevant concepts from distributed consensus averaging algorithm, graph symmetry and semidefinite programming. Sections III contains the proposed method and main results of the paper, namely the exact determination of optimal weights for fastest distribution consensus algorithm via stratification and SDP in TFS network. Section IV presents simulations demonstrating improvement of the obtained optimal weights over other weighting methods and tradeoff between the parameters of network and section V concludes the paper.

## II. PRELIMINARIES

This section introduces the notation used in the paper and reviews relevant concepts from distributed consensus averaging algorithm, graph symmetry and semidefinite programming.

*A. Distributed Consensus*

We consider a network $\mathcal{N}$ with the associated graph $\mathcal{G} = (\mathcal{V}, \mathcal{E})$ consisting of a set of nodes $\mathcal{V}$ and a set of edges $\mathcal{E}$ where each edge$\{i,j\} \in \mathcal{E}$ is an unordered pair of distinct nodes.

Each node $i$ holds an initial scalar value $x_i(0) \in \mathbf{R}$, and $x^T(0) = (x_1(0), \ldots, x_n(0))$ denotes the vector of initial node values on the network. Within the network two nodes can communicate with each other, if and only if they are neighbors.

The main purpose of distributed consensus averaging is to compute the average of the initial values, $(1/n) \sum_{i=1}^{n} x_i(0)$ via a distributed algorithm, in which the nodes only communicate with their neighbors. In this work, we consider distributed linear iterations, which have the form

$$x_i(t+1) = W_{ii} x_i(t) + \sum_{j \neq i} W_{ij} x_j(t), \quad i = 1, \ldots, n$$

where $t = 0, 1, 2, \ldots$ is the discrete time index and $W_{ij}$ is the weight on $x_j$ at node $i$ and the weight matrix have the same sparsity pattern as the adjacency matrix of the network's associated graph or $W_{ij} = 0$ if $\{i,j\} \notin \mathcal{E}$, this iteration can be written in vector form as

$$x(t+1) = Wx(t) \tag{1}$$

The linear iteration (1) implies that $x(t) = W^t x(0)$ for $= 0, 1, 2, \ldots$. We want to choose the weight matrix $W$ so that for any initial value $x(0)$, $x(t)$ converges to the average vector $\bar{x} = (\mathbf{1}^T x(0)/n)\mathbf{1} = (\mathbf{11}^T/n) x(0)$ i.e.,

$$\lim_{t \to \infty} x(t) = \lim_{t \to \infty} W^t x(0) = \frac{\mathbf{11}^T}{n} x(0) \tag{2}$$

(Here **1** denotes the column vector with all coefficients one). This is equivalent to the matrix equation

$$\lim_{t \to \infty} W^t = \frac{\mathbf{11}^T}{n} \tag{3}$$

Assuming (2-3) holds, the *convergence factor* can be defined as

$$r(W) = \sup \frac{\|x(t+1) - \bar{x}\|_2}{\|x(t) - \bar{x}\|_2}$$

where $\|\cdot\|_2$ denotes the spectral norm, or maximum singular value. The FDC problem in terms of the convergence factor can be expressed as the following optimization problem:

$$\begin{aligned} \min_{W} \quad & r(W) \\ s.t. \quad & \lim_{t \to \infty} W^t = \mathbf{1}\mathbf{1}^T/n, \\ & \forall \{i,j\} \notin \mathcal{E}: W_{ij} = 0 \end{aligned} \quad (4)$$

where $W$ is the optimization variable, and the network is the problem data.

In [5] it has been shown that the necessary and sufficient conditions for the matrix equation (3) to hold is that one is a simple eigenvalue of $\mathbf{W}$ associated with the eigenvector $\mathbf{1}$, and all other eigenvalues are strictly less that one in magnitude. Moreover in [5] FDC problem has been formulated as the following minimization problem

$$\begin{aligned} \min_{W} \quad & \max(\lambda_2, -\lambda_n) \\ s.t. \quad & W = W^T, W\mathbf{1} = \mathbf{1} \\ & \forall \{i,j\} \notin \mathcal{E}: W_{ij} = 0 \end{aligned} \quad (5)$$

Where $1 = \lambda_1 \geq \lambda_2 \geq \cdots \geq \lambda_n \geq -1$ are eigenvalues of $W$ arranged in decreasing order and $\max(\lambda_2, -\lambda_n)$ is the *Second Largest Eigenvalue Modulus* (*SLEM*) of $W$, and the main problem can be formulated in the semidefinite programming form as [5]:

$$\begin{aligned} \min_{W} \quad & s \\ s.t. \quad & -sI \preccurlyeq W - \mathbf{1}\mathbf{1}^T/n \preccurlyeq sI \\ & W = W^T, W\mathbf{1} = \mathbf{1} \\ & \forall \{i,j\} \notin \mathcal{E}: W_{ij} = 0 \end{aligned} \quad (6)$$

We refer to problem (6) as the Fastest Distributed Consensus (FDC) averaging problem.

*B. Symmetry of graphs*

An automorphism of a graph $\mathcal{G} = (\mathcal{V}, \mathcal{E})$ is a permutation $\sigma$ of $\mathcal{V}$ such that $\{i,j\} \in \mathcal{E}$ if and only if $\{\sigma(i), \sigma(j)\} \in \mathcal{E}$, the set of all such permutations, with composition as the group operation, is called the automorphism group of the graph and denoted by $Aut(\mathcal{G})$. For a vertex $i \in \mathcal{V}$, the set of all images $\sigma(i)$, as $\sigma$ varies through a subgroup $G \subseteq Aut(\mathcal{G})$, is called the orbit of $i$ under the action of $G$. The vertex set $\mathcal{V}$ can be written as disjoint union of distinct orbits. In [26], it has been shown that the weights on the edges within an orbit must be the same.

## C. Semidefinite Programming (SDP)

SDP is a particular type of convex optimization problem [27]. An SDP problem requires minimizing a linear function subject to a linear matrix inequality constraint [28]:

$$\begin{aligned} \min \quad & \rho = c^T x, \\ s.t. \quad & F(x) \geq 0 \end{aligned} \quad (7)$$

where $c$ is a given vector, $x^T = (x_1, \ldots, x_n)$, and $F(x) = F_0 + \sum_i x_i F_i$, for some fixed hermitian matrices $F_i$. The inequality sign in $F(x) \geq 0$ means that $F(x)$ is positive semi-definite.

This problem is called the primal problem. Vectors $x$ whose components are the variables of the problem and satisfy the constraint $F(x) \geq 0$ are called primal feasible points, and if they satisfy $F(x) \geq 0$, they are called strictly feasible points. The minimal objective value $c^T x$ is by convention denoted by $\rho^*$ and is called the primal optimal value.

Due to the convexity of the set of feasible points, SDP has a nice duality structure, with the associated dual program being:

$$\begin{aligned} \max \quad & -Tr[F_0 Z] \\ s.t. \quad & Z \geq 0 \\ & Tr[F_i Z] = c_i \end{aligned} \quad (8)$$

Here the variable is the real symmetric (or Hermitian) positive matrix $Z$, and the data $c$, $F_i$ are the same as in the primal problem. Correspondingly, matrix $Z$ satisfying the constraints is called dual feasible (or strictly dual feasible if $Z > 0$). The maximal objective value of $-Tr[F_0 Z]$, i.e. the dual optimal value is denoted by $d^*$.

The objective value of a primal (dual) feasible point is an upper (lower) bound on $\rho^*(d^*)$. The main reason why one is interested in the dual problem is that one can prove that $d^* \leq \rho^*$, and under relatively mild assumptions, we can have $\rho^* = d^*$. If the equality holds, one can prove the following optimality condition on $x$.

A primal feasible $x$ and a dual feasible $Z$ are optimal, which is denoted by $\hat{x}$ and $\hat{Z}$, if and only if

$$F(\hat{x})\hat{Z} = \hat{Z}F(\hat{x}) = 0. \quad (9)$$

This latter condition is called the complementary slackness condition.

In one way or another, numerical methods for solving SDP problems always exploit the inequality $d \leq d^* \leq \rho^* \leq \rho$, where $d$ and $\rho$ are the objective values for any dual feasible point and primal feasible point, respectively. The difference

$$\rho^* - d^* = c^T x + Tr[F_0 Z] = Tr[F(x)Z] \geq 0$$

is called the duality gap. If the equality $d^* = \rho^*$ holds, i.e. the optimal duality gap is zero, then we say that strong duality holds.

### III. TWO FUSED STAR (TFS) NETWORK

In this section we solve the Fastest Distributed Consensus (FDC) averaging algorithm for Two Fused Star (TFS) network consisting of two different symmetric star networks which share the same central node, by means of stratification and Semidefinite Programming (SDP).

*A. Two Fused Star (TFS) Network*

TFS Network consisting of path formed branches called tails with two different lengths, $m_1$ and $m_2$ where the numbers of branches are $n_1$ and $n_2$, respectively and the tails are connected to one node called central node, we call the whole network TFS network. (see Fig.1. for $n_1 = 2, m_1 = 3, n_2 = 3, m_2 = 2$). The connectivity graph of TFS network has $|\mathcal{V}| = m_1 n_1 + m_2 n_2 + 1$ nodes and $|\mathcal{E}| = m_1 n_1 + m_2 n_2$ edges, where the set of nodes is denoted by

$$\mathcal{V} = \{(-m_1, 1), (-m_1, 2), \dots, (-m_1, n_1), (-m_1 + 1, 1), \dots, (-1, n_1), (0,0), (1,1), (1,2), \dots, (1, n_2), (2,1), \dots, (m_2, n_2)\}.$$

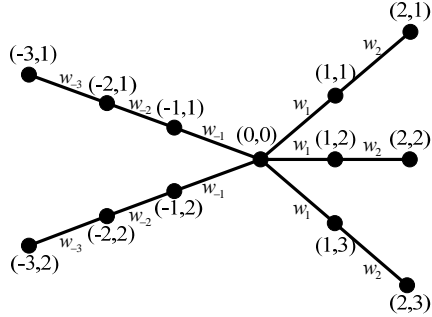

Fig.1. Stratums of weighted TFS for $n_1 = 2, m_1 = 3, n_2 = 3, m_2 = 2$.

Automorphism of TFS graph is $S_{n_1} \otimes S_{n_2}$ permutation of tails, hence according to subsection II-B it has $m_1 + m_2 + 1$ class of edge orbits, thus it suffices to consider just $m_1 + m_2$ weights $w_{-m_1}, \dots, w_{-1}, w_1, \dots, w_{m_2}$ (as labeled in Fig. 1. for $n_1 = 2, m_1 = 3, n_2 = 3, m_2 = 2$) and consequently the weight matrix is defined as

$$W_{(i,\mu),(j,\eta)} = \begin{cases} w_i & if \ \mu = \eta = 1, \dots, n_1, \ i = j - 1 = -m_1, \dots, -2 \\ w_{-1} & if \ i = -1, \ \mu = 1, \dots, n_1, \ j = \eta = 0 \\ w_1 & if \ i = \mu = 0, \ \eta = 1, \dots, n_2, \ j = 1 \\ w_i & if \ \mu = \eta = 1, \dots, n_2, \ i + 1 = j = 2, \dots, m_2 \\ 1 - w_{-m_1} & if \ i = j = -m_1, \ \mu = \eta = 1, \dots, n_1 \\ 1 - w_{i-1} - w_i & if \ i = j = -m_1 + 1, \dots, -1, \ \mu = \eta = 1, \dots, n_1 \\ 1 - n_1 w_{-1} - n_2 w_1 & if \ i = j = \mu = \eta = 0 \\ 1 - w_i - w_{i+1} & if \ i = j = 1, \dots, m_2 - 1, \ \mu = \eta = 1, \dots, n_2 \\ 1 - w_{m_2} & if \ i = j = m_2, \ \mu = \eta = 1, \dots, n_2 \\ 0 & otherwise \end{cases}$$

Introducing the orthonormal basis $e_{i,\mu} = e_i \otimes e_\mu$ for $\{i, \mu\} = \{i = -m_1, \dots, -1, \ \mu = 1, \dots, n_1\} \cup \{i = \mu = 0\} \cup \{i = 1, \dots, m_2, \ \mu = 1, \dots, n_2\}$ where $e_i$ and $e_\mu$ are $(m_1 + m_2 + 1) \times 1$ and $(n_1 + n_2) \times 1$ column vectors with one in the $i$-th and $\mu$-th position respectively and zero elsewhere, the weight matrix can be written as:

$$W = \sum_{i=-m_1}^{-2} \sum_{\mu=1}^{n_1} w_i e_{i,\mu} e_{i+1,\mu}^T + \sum_{\mu=1}^{n_1} w_{-1} e_{-1,\mu} e_{0,0}^T + \sum_{\eta=1}^{n_2} w_1 e_{0,0} e_{1,\eta}^T + \sum_{i=2}^{m_2} \sum_{\mu=1}^{n_2} w_i e_{i-1,\mu} e_{i,\mu}^T$$

$$+ \sum_{\mu=1}^{n_1} (1 - w_{-m_1}) e_{-m_1,\mu} e_{-m_1,\mu}^T + \sum_{i=-m_1+1}^{-1} \sum_{\mu=1}^{n_1} (1 - w_{i-1} - w_i) e_{i,\mu} e_{i,\mu}^T$$

$$+ \sum_{i=1}^{m_2-1} \sum_{\mu=1}^{n_2} (1 - w_i - w_{i+1}) e_{i,\mu} e_{i,\mu}^T + \sum_{\mu=1}^{n_2} (1 - w_{m_2}) e_{m_2,\mu} e_{m_2,\mu}^T + (1 - n_1 w_{-1} - n_2 w_1) e_{0,0} e_{0,0}^T$$

Denoting the $i$-th vertex orbit (called $i$-th stratum) under the $S_{n_1} \otimes S_{n_2}$ permutation by $\Gamma_i = \{(i, \mu) : (i, \mu) \in O(i, 1), i = -m_1, \dots, m_2\}$, the vertex set of TFS graph $\mathcal{V}$ can be written as the disjoint union of strata $\Gamma_i$ as

$$\mathcal{V} = \bigcup_{\mu=0}^{m} \Gamma_\mu$$

### B. Stratification of TFS Network

Using Stratification method introduced in [26, 29, 30, 31, 32, 33], the TFS graph can be stratified into a disjoint union of $m_1 + m_2 + 1$ strata as shown in Fig. 1. for $n_1 = 3, m_1 = 2, n_2 = 2, m_2 = 3$.

In each strata (except for $\Gamma_0$), the unitary DFT matrices of size $(m_1 + m_2) \times (m_1 + m_2)$ and $(m_1 + m_2 + 1) \times (m_1 + m_2 + 1)$ map set of orthonormal vectors in strata to a new set of orthonormal vectors defined as

$$\varphi_{i,\mu} = \begin{cases} \dfrac{1}{\sqrt{n_1}} \sum_{k=1}^{n_1} \omega_1^{\mu k} e_{i,k} & i = -m_1, \dots, -1 \\ & \mu = 0, \dots, n_1 - 1 \\ e_{0,0} & i = \mu = 0 \\ \dfrac{1}{\sqrt{n_2}} \sum_{k=1}^{n_2} \omega_2^{\mu k} e_{i,k} & i = 1, \dots, m_2 \\ & \mu = 0, \dots, n_2 - 1 \end{cases}$$

where $\omega_1 = e^{j\frac{2\pi}{n_1}}$ and $\omega_2 = e^{j\frac{2\pi}{n_2}}$. Considering new basis $\{\varphi_{i,\mu}|i=-m_1,\ldots,-1;\mu=0,\ldots,n_1-1\} \cup \{\varphi_{0,0}\} \cup \{\varphi_{i,\mu}|i=1,\ldots,m_2;\mu=0,\ldots,n_2-1\}$, the weight matrix $W$ has the matrix elements in the new basis as provided in Appendix A. Therefore the weight matrix $W$ has the following block diagonal form in the new basis.

$$W = \begin{bmatrix} W_{-1} & 0 & \cdots & & & & & 0 \\ 0 & \ddots & 0 & \cdots & & & & \\ \vdots & 0 & W_{-1} & 0 & \cdots & & & \\ & \vdots & 0 & W_0 & 0 & \vdots & & \\ & & \vdots & 0 & W_1 & 0 & \vdots & \\ & & & \cdots & 0 & \ddots & 0 & \\ 0 & & & & & \cdots & 0 & W_1 \end{bmatrix}$$

where $W_{-1}, W_0, W_1$ are provided in Appendix A. The eigenvalues of $W$ can be obtained from diagonalization of the matrices, $W_{-1}$, $W_0$ and $W_1$. Introducing $W'$ as

$$W' = \begin{bmatrix} W_{-1} & 0 \\ 0 & W_1 \end{bmatrix} \tag{10}$$

while considering the fact that $W'$ is a submatrix of $W_0$ and using *Cauchy Interlacing Theorem*,

*Theorem 1* (*Cauchy Interlacing Theorem*) [34]:

Let $A$ and $B$ be $n \times n$ and $m \times m$ matrices, where $m \leq n$, $B$ is called a compression of $A$ if there exists an orthogonal projection $P$ onto a subspace of dimension $m$ such that $PAP = B$. The Cauchy interlacing theorem states that If the eigenvalues of $A$ are $\lambda_1(A) \leq \cdots \leq \lambda_n(A)$, and those of $B$ are $\lambda_1(B) \leq \cdots \leq \lambda_m(B)$, then for all $j$,

$$\lambda_j(A) \leq \lambda_j(B) \leq \lambda_{n-m+j}(A)$$

Notice that, when $n - m = 1$, we have

$$\lambda_j(A) \leq \lambda_j(B) \leq \lambda_{j+1}(A)$$

we can state the following corollary for the eigenvalues of $W_0$ and $W'$.

In the case of $n_1 = 1$, the weight matrix $W$ does not include $W_{-1}$ and $W'$ reduces to $W_1$ and consequently difference between dimensions of $W_0$ and $W'$ will be more than one and Cauchy interlacing theorem will not be true. It is clear that the same result holds for $n_2 = 1$ and $W_1$, thus the followings are true for $n_1, n_2 \geq 2$.

*Corollary 1*,

If we consider $W_0$ and $W'$ given in (A-1) and (10) respectively, then theorem 1 implies the following relations between the eigenvalues of $W_0$ and $W'$

$$\lambda_{m_1+m_2+1}(W_0) \leq \lambda_{m_1+m_2}(W') \leq \cdots \leq \lambda_2(W') \leq \lambda_2(W_0) \leq \lambda_1(W') \leq \lambda_1(W_0) = 1$$

It is obvious from this result that the eigenvalues $\lambda_2(W)$ and $\lambda_{m_1+m_2+1}(W)$ are amongst the eigenvalues of $W'$ and $W_0$, respectively.

*C. Determination of Optimal Weights for FDC Algorithm in TFS Network via SDP*

Based on the corollary 2 and subsection II-A, one can express FDC problem for TFS network in the form of semidefinite programming as:

$$\begin{aligned} \min \quad & s \\ s.t. \quad & W' \leq sI \\ & -sI \leq W_0 - \boldsymbol{v}\boldsymbol{v}^T \end{aligned} \quad (11)$$

where $\boldsymbol{v}$ is a $(m_1 + m_2 + 1) \times 1$ column vector defined as:

$$v_i = \frac{1}{\sqrt{m_1 n_1 + 1 + m_2 n_2}} \times \begin{cases} \sqrt{n_1} & \text{for } i = 1, \ldots, m_1 \\ 1 & \text{for } i = m_1 + 1 \\ \sqrt{n_2} & \text{for } i = m_1 + 2, \ldots, m_1 + m_2 + 1 \end{cases} \quad (12)$$

which is eigenvector of $W_0$ corresponding to the eigenvalue one.

$W_0$ and $W'$ can be written as a linear combination of rank one matrices,

$$W_0 = I - \sum_{i=-m_1}^{-2} 2w_i \boldsymbol{\alpha}_i \boldsymbol{\alpha}_i^T - \sum_{i=2}^{m_2} 2w_i \boldsymbol{\alpha}_i \boldsymbol{\alpha}_i^T - (n_1+1)w_{-1}\boldsymbol{\alpha}_{-1}\boldsymbol{\alpha}_{-1}^T - (n_2+1)w_1\boldsymbol{\alpha}_1\boldsymbol{\alpha}_1^T \quad (13)$$

$$W' = I - \sum_{i=-m_1}^{-2} 2w_i \boldsymbol{\alpha}'_i \boldsymbol{\alpha}'^T_i - \sum_{i=2}^{m_2} 2w_i \boldsymbol{\alpha}'_i \boldsymbol{\alpha}'^T_i - w_{-1}\boldsymbol{\alpha}'_{-1}\boldsymbol{\alpha}'^T_{-1} - w_1\boldsymbol{\alpha}'_1\boldsymbol{\alpha}'^T_1 \quad (14)$$

where for $i = -m_1, \ldots, m_2$ the vectors $\boldsymbol{\alpha}_i$ and $\boldsymbol{\alpha}'_i$ are $(m_1 + m_2 + 1) \times 1$ and $(m_1 + m_2) \times 1$ column vectors, respectively, as provided in Appendix B. Using (13) and (14), the constraints in (11) can be written as

$$sI - I + \sum_{i=-m_1}^{-2} 2w_i \boldsymbol{\alpha}'_i \boldsymbol{\alpha}_i^T + \sum_{i=2}^{m_2} 2w_i \boldsymbol{\alpha}'_i \boldsymbol{\alpha}_i^T - w_{-1}\boldsymbol{\alpha}'_{-1}\boldsymbol{\alpha}_{-1}^T + w_1\boldsymbol{\alpha}'_1\boldsymbol{\alpha}_1^T \geq 0 \quad (15\text{-a})$$

$$sI + I - \sum_{i=-m_1}^{-2} 2w_i \boldsymbol{\alpha}_i \boldsymbol{\alpha}_i^T - \sum_{i=2}^{m_2} 2w_i \boldsymbol{\alpha}_i \boldsymbol{\alpha}_i^T - (n_1+1)w_{-1}\boldsymbol{\alpha}_{-1}\boldsymbol{\alpha}_{-1}^T - (n_2+1)w_1\boldsymbol{\alpha}_1\boldsymbol{\alpha}_1^T - \boldsymbol{v}\boldsymbol{v}^T \geq 0 \quad (15\text{-b})$$

In order to formulate problem (11) in the form of standard semidefinite programming described in section II-C, we define $F_i, c_i$ and $x$ as below:

$$F_0 = \begin{bmatrix} I_{(m_1+m_2+1)\times(m_1+m_2+1)} - \boldsymbol{v}\boldsymbol{v}^T/n & 0 \\ 0 & -I_{(m_1+m_2)\times(m_1+m_2)} \end{bmatrix}$$

$$F_i = \begin{bmatrix} -\boldsymbol{\alpha}_i \boldsymbol{\alpha}_i^T & 0 \\ 0 & \boldsymbol{\alpha}'_i \boldsymbol{\alpha}_i'^T \end{bmatrix} \quad \text{for } i = -m_1, \ldots, m_2, \; i \neq -1,0,1$$

$$F_{-1} = \begin{bmatrix} -(n_1+1)\boldsymbol{\alpha}_{-1}\boldsymbol{\alpha}_{-1}^T & 0 \\ 0 & \boldsymbol{\alpha}'_{-1}\boldsymbol{\alpha}_{-1}'^T \end{bmatrix},$$

$$F_1 = \begin{bmatrix} -(n_2+1)\boldsymbol{\alpha}_1\boldsymbol{\alpha}_1^T & 0 \\ 0 & \boldsymbol{\alpha}'_1\boldsymbol{\alpha}_1'^T \end{bmatrix},$$

$$F_{-m_1-1} = I_{2m_1+2m_2+1}$$

$$c_{-m_1-1} = 1, \quad c_i = 0, \; i = -m_1, \ldots, m_2, \; i \neq 0$$

$$x^T = [x_{-m_1-1}, x_{-m_1}, \ldots, x_{-1}, x_1, \ldots, x_{m_2}] = [s, 2w_{-m_1}, \ldots, 2w_{-2}, w_{-1}, w_1, 2w_2, \ldots, 2w_{m_2}]$$

and in the dual case we choose the dual variable $Z \geq 0$ as

$$Z = \begin{bmatrix} Z_1 \\ Z_2 \end{bmatrix} \cdot [z_1^T \quad z_2^T] \quad (16)$$

where $z_1$ and $z_2$ are column vectors with $(m_1+m_2+1)$ and $m_1+m_2$ elements, respectively. Obviously (16) choice of $Z$ implies that it is positive definite.

From the complementary slackness condition (9) we have

$$(sI + W_0 - vv^T)z_1 = 0 \tag{17-a}$$

$$(sI - W')z_2 = 0 \tag{17-b}$$

Multiplying both sides of equation (17-a) by $vv^T$ we have $s(vv^T z_1) = 0$ which implies that

$$v^T z_1 = 0 \tag{18}$$

Using the constraints $Tr[F_i Z] = c_i$ we have

$$(\alpha_i^T z_1)^2 = (\alpha_i'^T z_2)^2, \quad i = -m_1, \dots, m_2, \quad i \neq -1, 1 \tag{19-a}$$

$$(n_1 + 1)(\alpha_{-1}^T z_1)^2 = (\alpha_{-1}'^T z_2)^2 \tag{19-b}$$

$$(n_2 + 1)(\alpha_1^T z_1)^2 = (\alpha_1'^T z_2)^2 \tag{19-c}$$

$$z_1^T z_1 + z_2^T z_2 = 1 \tag{20}$$

To have the strong duality we set $c^T x + Tr[F_0 Z] = 0$, hence we have

$$z_2^T z_2 - z_1^T z_1 = s \tag{21}$$

Considering the linear independence of $\alpha_i$ and $\alpha_i'$ for $i = -m_1, \dots, m_2$, we can expand $z_1$ and $z_2$ in terms of $\alpha_i$ and $\alpha_i'$ as

$$z_1 = \sum_{i=-m_1}^{m_2} a_i \alpha_i \tag{21-a}$$

$$z_2 = \sum_{i=-m_1}^{m_2} a_i' \alpha_i' \tag{21-b}$$

with the coordinates $a_i$ and $a_i'$, $i = -m_1, \dots, m_2$ to be determined.

Using (13) and (14) and the expansions (21), while considering (18), the slackness conditions (17), can be written as

$$(s + 1)a_i = 2w_i \alpha_i^T z_1, \tag{22-a}$$

$$(s+1)a_{-1} = (n_1+1)w_{-1}\boldsymbol{\alpha}_{-1}^T z_1, \tag{22-b}$$

$$(s+1)a_1 = (n_2+1)w_1\boldsymbol{\alpha}_1^T z_1, \tag{22-c}$$

$$(-s+1)a_i' = 2w_i\boldsymbol{\alpha}_i'^T z_2, \tag{23-a}$$

$$(-s+1)a_{-1}' = w_{-1}\boldsymbol{\alpha}_{-1}'^T z_2, \tag{23-b}$$

$$(-s+1)a_1' = w_1\boldsymbol{\alpha}_1'^T z_2, \tag{23-c}$$

where (22-a) and (23-a) hold for $i = -m_1, \ldots, m_2$ and $i \neq -1, 1$. Considering (19), (22) and (23), we obtain

$$(s+1)^2 a_i^2 = (-s+1)^2 a_i'^2, \tag{24}$$

for $i = -m_1, \ldots, m_2$, or equivalently

$$\frac{a_i^2}{a_j^2} = \frac{a_i'^2}{a_j'^2} \tag{25}$$

for $\forall i, j = [-m_1, m_2]$ and for $\boldsymbol{\alpha}_i^T z_1$ and $\boldsymbol{\alpha}_i'^T z_2$, we have

$$\boldsymbol{\alpha}_i^T z_1 = \sum_{j=1}^{n-1} a_j G_{i,j} \tag{26-a}$$

$$\boldsymbol{\alpha}_i'^T z_2 = \sum_{j=1}^{n-1} a_j' G_{i,j}' \tag{26-b}$$

where $G$ and $G'$ are the Gram matrices, defined as

$$G_{i,j} = \boldsymbol{\alpha}_i^T \boldsymbol{\alpha}_j$$

$$G_{i,j}' = \boldsymbol{\alpha}_i'^T \boldsymbol{\alpha}_i'$$

with $G$ and $G'$ as provided in Appendix B. Substituting (26) in (23) we have

$$(s + 1 - 2w_{-m_1})a_{-m_1} = -w_{-m_1}a_{-m_1+1} \quad (27\text{-a})$$

$$(s + 1 - 2w_i)a_i = -w_i(a_{i-1} + a_{i+1}) \quad (27\text{-b})$$

$$(s + 1 - 2w_{-2})a_{-2} = -w_{-2}(a_{-3} + \hat{a}_{-1}) \quad (27\text{-c})$$

$$(s + 1 - (n_1 + 1)w_{-1})\hat{a}_{-1} = -w_{-1}a_{-2} - \sqrt{n_1 n_2}w_{-1}\hat{a}_1 \quad (27\text{-d})$$

$$(s + 1 - (n_2 + 1)w_1)\hat{a}_1 = -\sqrt{n_1 n_2}w_1\hat{a}_{-1} - w_1 a_2 \quad (27\text{-e})$$

$$(s + 1 - 2w_2)a_2 = -w_2(\hat{a}_1 + a_3) \quad (27\text{-f})$$

$$(s + 1 - 2w_{m_2})a_{m_2} = -w_{m_2}a_{m_2-1} \quad (27\text{-g})$$

and

$$(-s + 1 - 2w_{-m_1})a'_{-m_1} = -w_{-m_1}a'_{-m_1+1} \quad (28\text{-a})$$

$$(-s + 1 - 2w_i)a'_i = -w_i(a'_{i-1} + a'_{i+1}) \quad (28\text{-b})$$

$$(-s + 1 - 2w_{-2})a'_{-2} = -w_{-2}(a'_{-3} + \hat{a}'_{-1}) \quad (28\text{-c})$$

$$(-s + 1 - w_{-1})\hat{a}'_{-1} = -w_{-1}a'_{-2} \quad (28\text{-d})$$

$$(-s + 1 - w_1)\hat{a}'_1 = -w_1 a'_2 \quad (28\text{-e})$$

$$(-s + 1 - 2w_2)a'_2 = -w_2(\hat{a}'_1 + a'_3) \quad (28\text{-f})$$

$$(-s + 1 - 2w_{m_2})a'_{m_2} = -w_{m_2}a'_{m_2-1} \quad (28\text{-g})$$

where $\hat{a}_{-1} = \left(\sqrt{2}/\sqrt{(n_1 + 1)}\right)a_{-1}$, $\hat{a}'_{-1} = -\sqrt{2}a'_{-1}$, $\hat{a}_1 = \left(\sqrt{2}/\sqrt{(n_2 + 1)}\right)a_1$, $\hat{a}'_1 = \sqrt{2}a'_1$ and (27-b) and (28-b) hold for $i = -m_1 + 1, \ldots, m_2 - 1,$, $i \neq -2, -1, 1, 2$.

Now we can determine $s$ (*SLEM*), the optimal weights and the coordinates $a_i$ and $a'_i$, in an inductive manner as follows:

In the first stage, from comparing equations (27-a) and (28-a) and considering the relation (25), we can conclude that

$$(-s + 1 - 2w_{-m_1})^2 = (s + 1 - 2w_{-m_1})^2 \quad (29)$$

which results in $w_{-m_1} = 1/2$ and $s = 0$, where the latter is not acceptable. Assuming $s = \cos(\theta)$ and substituting $w_{-m_1} = 1/2$ in (27-a) and (28-a), we have

$$a_{-m_1+1} = \frac{\sin(2(\pi - \theta))}{\sin(\pi - \theta)} a_{-m_1} \tag{30-a}$$

$$a'_{-m_1+1} = \frac{\sin(2\theta)}{\sin(\theta)} a'_{-m_1} \tag{30-b}$$

Continuing the above procedure inductively, up to $i - 1$ stages, and assuming

$$a_j = \frac{\sin((j + 1 + m_1)(\pi - \theta))}{\sin(\pi - \theta)} a_{-m_1}, \qquad \forall j \leq i \leq -1$$

and

$$a'_j = \frac{\sin((j + 1 + m_1)\theta)}{\sin(\theta)} a'_{-m_1} \qquad \forall j \leq i \leq -1$$

for the $i$-th stage, we get the following equations from comparison of equations (27-b) and (28-b),

$$\left((s + 1 - 2w_i)\frac{\sin((i + 1 + m_1)(\pi - \theta))}{\sin(\pi - \theta)} + w_i \frac{\sin((i + m_1)(\pi - \theta))}{\sin(\pi - \theta)}\right) a_{-m_1} = -w_i a_{i+1} \tag{31-a}$$

$$\left((-s + 1 - 2w_i)\frac{\sin((i + 1 + m_1)\theta)}{\sin(\theta)} + w_i \frac{\sin((i + m_1)\theta)}{\sin(\theta)}\right) a'_{-m_1} = -w_i a'_{i+1} \tag{31-b}$$

while considering relation (25) we can conclude that

$$\left((-s + 1 - 2w_i)\sin((i + 1 + m_1)\theta) + w_i \sin((i + m_1)\theta)\right)^2$$
$$= \left((s + 1 - 2w_i)\sin((i + 1 + m_1)(\pi - \theta)) + w_i \sin((i + m_1)(\pi - \theta))\right)^2$$

which results in

$$w_i = \frac{1}{2} \tag{32}$$

Substituting $w_i = 1/2$ in (31), we have

$$a_{i+1} = \frac{\sin\big((i+2+m_1)(\pi-\theta)\big)}{\sin(\pi-\theta)} a_{-m_1} \tag{33-a}$$

$$a'_{i+1} = \frac{\sin\big((i+2+m_1)\theta\big)}{\sin\theta} a'_{-m_1} \tag{33-b}$$

Since the equations (27-b) and (28-b) does not hold for $i = -2$, the results in (32) and (33) are true for $i = -m_1, \dots, -3$,

and in the $(m_1 - 2)$-th stage, from comparing equations (27-c) and (28-c) and considering the relation (25), we can conclude that

$$w_{-2} = \frac{1}{2} \tag{34}$$

$$\hat{a}_{-1} = \frac{\sin\big(m_1(\pi-\theta)\big)}{\sin(\pi-\theta)} a_{-m_1} \tag{35-a}$$

$$\hat{a}'_{-1} = \frac{\sin(m_1\theta)}{\sin\theta} a'_{-m_1} \tag{35-b}$$

The same inductive procedure can be used to obtain the weights with positive indices, simply by using equations (27-b), (28-b), (27-f), (28-f), (27-g) and (28-g) and relation (25), which results in

$$w_i = \frac{1}{2} \tag{36}$$

$$a_{i-1} = \frac{\sin\big((m_2-i+2)(\pi-\theta)\big)}{\sin(\pi-\theta)} a_{m_2} \tag{37-a}$$

$$\hat{a}_1 = \frac{\sin\big((m_2)(\pi-\theta)\big)}{\sin(\pi-\theta)} a_{m_2} \tag{37-b}$$

$$a'_{i-1} = \frac{\sin\big((m_2-i+2)\theta\big)}{\sin(\theta)} a'_{m_2} \tag{37-c}$$

$$\hat{a}'_1 = \frac{\sin(m_2\theta)}{\sin(\theta)} a'_{m_2} \tag{37-d}$$

where (36) and (37-a) and (37-c) are true for $i = 3, \dots, m_2$.

Using equations (33), (35) and (37) we can express $a_{-2}, \hat{a}_{-1}, a'_{-2}, \hat{a}'_{-1}$ and $\hat{a}_1, a_2, \hat{a}'_1, a'_2$ in terms of $a_{-m_1}, a'_{-m_1}, a_{m_2}, a'_{m_2}$, and substituting the results in equations (27-d), (27-e), (28-d), and (28-e) we have:

$$\left((s+1-(n_1+1)w_{-1})\sin(m_1(\pi-\theta))+w_{-1}\sin((m_1-1)(\pi-\theta))\right)a_{-m_1}$$
$$=-\sqrt{n_1 n_2}w_{-1}\sin(m_2(\pi-\theta))a_{m_2} \qquad (38\text{-a})$$

$$\left((s+1-(n_2+1)w_1)\sin(m_2(\pi-\theta))+w_1\sin((m_2-1)(\pi-\theta))\right)a_{m_2}$$
$$=-\sqrt{n_1 n_2}w_1\sin(m_1(\pi-\theta))a_{-m_1} \qquad (38\text{-b})$$

$$(-s+1-w_{-1})\sin(m_1\theta)=-w_{-1}\sin((m_1-1)\theta) \qquad (38\text{-c})$$

$$(-s+1-w_1)\sin(m_2\theta)=-w_1\sin((m_2-1)\theta) \qquad (38\text{-d})$$

from (38-c) and (38-d) we can conclude that

$$w_{-1}=\frac{(1-s)\sin(m_1\theta)}{\sin(m_1\theta)-\sin((m_1-1)\theta)} \qquad (39\text{-a})$$

$$w_1=\frac{(1-s)\sin(m_2\theta)}{\sin(m_2\theta)-\sin((m_2-1)\theta)} \qquad (39\text{-b})$$

Substituting (39) in (38-a) and (38-b), we obtain

$$\frac{a_{m_2}}{a_{-m_1}}=\frac{2s(\sin(m_1\theta)-\sin((m_1-1)\theta))+n_1(s-1)\sin(m_1\theta)}{(s-1)\sqrt{n_1 n_2}\sin(m_2\theta)} \qquad (40\text{-a})$$

$$\frac{a_{-m_1}}{a_{m_2}}=\frac{2s(\sin(m_2\theta)-\sin((m_2-1)\theta))+n_2(s-1)\sin(m_2\theta)}{(s-1)\sqrt{n_1 n_2}\sin(m_1\theta)} \qquad (40\text{-b})$$

where by substituting $s=\cos(\theta)$ in (40), we can conclude that $\theta$ has to satisfy the following relation

$$\left(\frac{2}{n_1}\cot(m_1\theta)\cot(\theta/2)-1\right)\times\left(\frac{2}{n_2}\cot(m_2\theta)\cot(\theta/2)-1\right)=1 \qquad (41)$$

In the case of $m_1=m_2=m$, (symmetric star) equation (41) reduces to

$$(n_1+n_2+2)\times\cos\left(\left(m+\frac{1}{2}\right)\theta\right)=(n_1+n_2-2)\times\cos\left(\left(m-\frac{1}{2}\right)\theta\right) \qquad (42)$$

which is in agreement with the results of [24].

Also one should notice that necessary and sufficient conditions for the convergence of weight matrix are satisfied, since all roots of $s$ which are the eigenvalues of $W$ are strictly less than one in magnitude, and one is a simple eigenvalue of $W$ associated with the eigenvector **1**, to support this fact we have computed numerically the roots of equation (41) whereby considering the relation $s = \cos(\theta)$ and that all roots of (41) are simple, we can conclude that all roots of $s$ are strictly less than one in magnitude and in addition the smallest and second largest roots of $s$ are listed in Table .1. for different values of $m_1, n_1, m_2$ and $n_2$.

| $m_1, n_1, m_2, n_2$ | 2$^{nd}$ Largest Eigenvalue | Smallest Eigenvalue |
|---|---|---|
| (3,4,4,3) | 0.9545 | -0.9445 |
| (10,20,20,10) | 0.997739 | -0.997739 |
| (100,200,200,100) | 0.9999772 | -0.9999772 |

Table. 1. Second Largest Eigenvalue and Smallest Eigenvalue of TFS network for different values of $m_1, n_1, m_2$ and $n_2$

As it is obvious from the results depicted in Table. 1. the *SLEM* of TFS network increases with the length of branches of network which is due to the topology of TFS network and in the case of optimum weights, the second largest eigenvalue and the smallest eigenvalue have the same absolute values.

## IV. SIMULATION RESULTS

The aim of this section is to show the improvement of optimal weights obtained in section III over other weighting methods, namely maximum degree, Metropolis and best constant weighting methods by evaluating *SLEM* numerically for different weighting methods, moreover we have investigated the tradeoff between the parameters of network and convergence rate by numerical results.

In Table. 2. *SLEM* of TFS network for optimal weights, Maximum degree, Metropolis and best constant weighting methods has been depicted for different sizes of TFS network.

| $m_1, n_1, m_2, n_2$ | Optimal Weights | Max Degree | Metropolis | Best Constant |
|---|---|---|---|---|
| (3,4,4,3) | 0.95450 | 0.98277 | 0.97194 | 0.97089 |
| (3,4,3,6) | 0.95381 | 0.98019 | 0.97195 | 0.96497 |
| (10,20,20,10) | 0.99774 | 0.99981 | 0.99884 | 0.99962 |

Table. 2. *SLEM* of TFS network for optimal weights, maximum degree, Metropolis and best constant weighting methods.

Now we compare a TFS network with its equivalent Symmetric Star network. To do so, we define the total number of branches $n$ and the average length of branches $\bar{m}$, of the equivalent symmetric star, in term of parameters of TFS network as follows:

$$n = n_1 + n_2,$$

$$\bar{m} = \frac{m_1 n_1 + m_2 n_2}{n_1 + n_2}. \tag{43}$$

where $SLEM = \max_\theta |\cos(\theta)|$ and $\theta$ is obtained from numerical solution of (41) and (42) for TFS network and its equivalent symmetric star network, respectively. In Fig. 2. *SLEM* of TFS network and its equivalent Symmetric Star network are depicted in terms of the average length of branches $\bar{m}$, for $n_1 = 6$ and $n_2 = 12$.

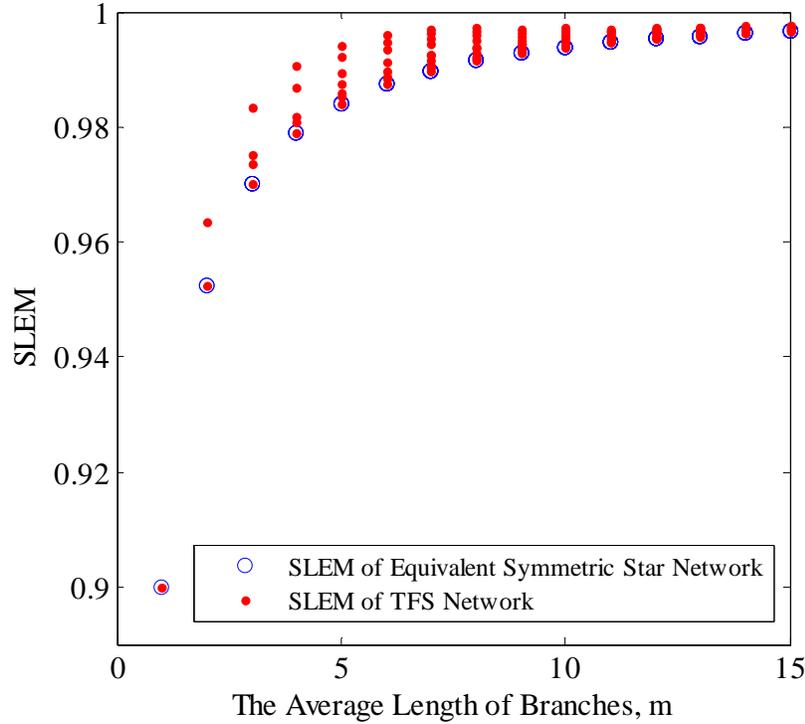

Fig.2. *SLEM* of TFS network and its equivalent Symmetric Star network in terms of integer $\bar{m}$ for $n_1 = 6$ and $n_2 = 12$.

As it is clear from Fig. 2. for all values of $\bar{m}$, SLEM of the equivalent Symmetric Star network is smaller than SLEM of other TFS networks with the same average length of branches, which in turn means that the equivalent Symmetric Star network converges faster than the other TFS networks with the same average length of branches.

In Fig. 3. SLEM of TFS network is depicted in terms of the length of branches $m_1$ and $m_2$, for $n_1 = 2$ and $n_2 = 22$.

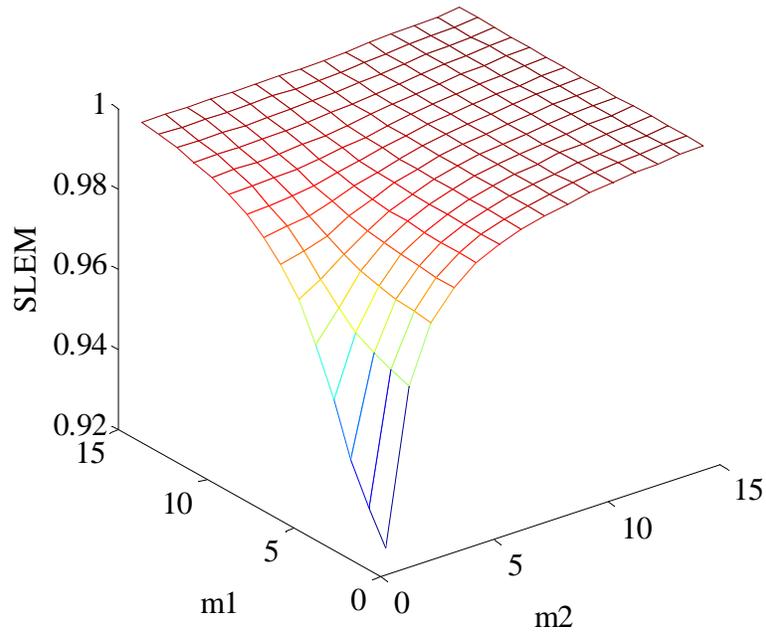

Fig.3. *SLEM* of TFS network in terms of the length of branches $m_1$ and $m_2$, for $n_1 = 2$ and $n_2 = 22$.

It is obvious from Fig. 3. that SLEM increases with $m_2$ faster than $m_1$.

In Fig. 4. $w_{-1}$ the weight of edges which are connecting branches of first star of TFS network to the central node is depicted in terms of the length of branches $m_1$ and $m_2$, for $n_1 = 2$ and $n_2 = 22$.

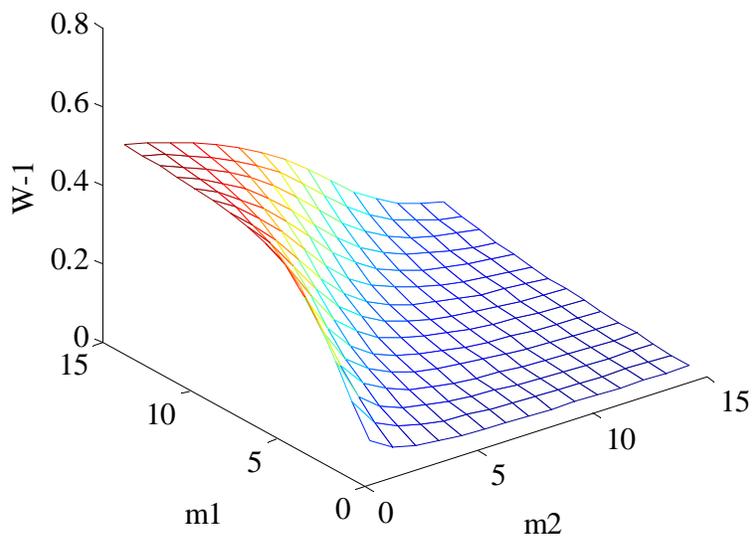

Fig.4. $w_{-1}$ in terms of $m_1$ and $m_2$, for $n_1 = 2$ and $n_2 = 22$.

It is obvious from Fig. 4. that $w_{-1}$ increases with $m_1$ while decreases with $m_2$, also it is interesting that the critical line in the curve of Fig. 4. is the average length of branches $\bar{m}$ defined in (43).

## V. CONCLUSION

Fastest Distributed Consensus averaging Algorithm in sensor networks has received renewed interest recently, but Most of the methods proposed so far usually avoid the direct computation of optimal weights and deal with the Fastest Distributed Consensus problem by numerical convex optimization methods.

Here in this work, we have solved Fastest Distributed Consensus problem for TFS network by means of stratification and Semidefinite Programming (SDP). Our approach is based on fulfilling the slackness conditions, where the optimal weights are obtained by inductive comparing of the characteristic polynomials initiated by slackness conditions. The simulation results confirm that the distributed consensus algorithm with optimal weights converges substantially faster than the one with other simple weighting methods, namely maximum degree, Metropolis and best constant weighting methods; moreover we have investigated the tradeoff between the parameters of network and convergence rate by numerical results. We believe that this method is powerful and lucid enough to be extended to other networks with more general topologies, namely star networks with more different types of branches, which is the object of future investigations.

## APPENDIX A

### ELEMENTS OF WEIGHT MATRIX IN THE BASIS DEFINED VIA STRATIFICATION

$W \cdot \varphi_{i,\mu}$

$$= \begin{cases} (1 - w_{-m_1})\varphi_{-m_1,\mu} + w_{-m_1}\varphi_{-m_1+1,\mu} & \text{for } i = -m_1; \ \mu = 0, \ldots, n_1 - 1 \\ w_{i-1}\varphi_{i-1,\mu} + (1 - w_{i-1} - w_i)\varphi_{i,\mu} + w_i\varphi_{i+1,\mu} & \text{for } i = -m_1 + 1, \ldots, -2; \ \mu = 0, \ldots, n_1 - 1 \\ w_{-2}\varphi_{-2,\mu} + (1 - w_{-2} - w_{-1})\varphi_{-1,\mu} & \text{for } i = -1; \ \mu = 1, \ldots, n_1 - 1 \\ w_{-2}\varphi_{-2,0} + (1 - w_{-2} - w_{-1})\varphi_{-1,0} + \sqrt{n_1}w_{-1}\varphi_{0,0} & \text{for } i = -1; \ \mu = 0 \\ \sqrt{n_1}w_{-1}\varphi_{-1,0} + (1 - n_1w_{-1} - n_2w_1)\varphi_{0,0} + \sqrt{n_2}w_1\varphi_{1,0} & \text{for } i = \mu = 0 \\ \sqrt{n_2}w_1\varphi_{0,0} + (1 - w_1 - w_2)\varphi_{1,0} + w_2\varphi_{2,0} & \text{for } i = 1; \ \mu = 0 \\ (1 - w_1 - w_2)\varphi_{1,\mu} + w_2\varphi_{2,\mu} & \text{for } i = 1; \ \mu = 1, \ldots, n_2 - 1 \\ w_i\varphi_{i-1,\mu} + (1 - w_i - w_{i+1})\varphi_{i,\mu} + w_{i+1}\varphi_{i+1,\mu} & \text{for } i = 2, \ldots, m_2; \ \mu = 0, \ldots, n_2 - 1 \\ w_{m_2}\varphi_{m_2-1,\mu} + (1 - w_{m_2})\varphi_{m_2,\mu} & \text{for } i = m_2; \ \mu = 0, \ldots, n_1 - 1 \end{cases}$$

$$W_{-1} = \begin{bmatrix} 1 - w_{-m_1} & w_{-m_1} & 0 & \cdots & 0 \\ w_{-m_1} & 1 - w_{-m_1} - w_{-m_1+1} & w_{-m_1+1} & \cdots & \vdots \\ 0 & w_{-m_1+1} & \ddots & \ddots & 0 \\ \vdots & \vdots & \ddots & \ddots & w_{-2} \\ 0 & \cdots & 0 & w_{-2} & 1 - w_{-2} - w_{-1} \end{bmatrix}$$

$$W_1 = \begin{bmatrix} 1-w_1-w_2 & w_2 & 0 & \cdots & 0 \\ w_2 & 1-w_2-w_3 & w_3 & \cdots & \vdots \\ 0 & w_3 & 1-w_3-w_4 & \cdots & 0 \\ \vdots & \vdots & \vdots & \ddots & w_{m_2} \\ 0 & \cdots & 0 & w_{m_2} & 1-w_{m_2} \end{bmatrix}$$

$$W_0 = \begin{bmatrix} W_{-1} & \sqrt{n_1}w_{-1} & 0 \\ \sqrt{n_1}w_{-1} & 1-n_1 w_{-1}-n_2 w_1 & \sqrt{n_2}w_1 \\ 0 & \sqrt{n_2}w_1 & W_1 \end{bmatrix} \tag{A-1}$$

# APPENDIX B

DEFINITION OF VECTORS $\boldsymbol{\alpha}_i$ AND $\boldsymbol{\alpha}'_i$ AND THEIR CORRESPONDING GRAM MATRICES $G$ AND $G'$

For $i = -m_1, \dots, m_2$ the vectors $\boldsymbol{\alpha}_i$ and $\boldsymbol{\alpha}'_i$ are defined as:

for $i = -m_1, \dots, -2$

$$\boldsymbol{\alpha}_{i,j} = \begin{cases} -1/\sqrt{2} & \text{for } j = i+m_1+1 \\ 1/\sqrt{2} & \text{for } j = i+m_1+2 \\ 0 & \text{Otherwise} \end{cases} \qquad \boldsymbol{\alpha}'_{i,j} = \begin{cases} -1/\sqrt{2} & \text{for } j = i+m_1+1 \\ 1/\sqrt{2} & \text{for } j = i+m_1+2 \\ 0 & \text{Otherwise} \end{cases}$$

for $i = -1$

$$\boldsymbol{\alpha}_{-1,j} = \frac{1}{\sqrt{n_1+1}} \times \begin{cases} -1 & \text{for } j = m_1 \\ \sqrt{n_1} & \text{for } j = m_1+1 \\ 0 & \text{Otherwise} \end{cases} \qquad \boldsymbol{\alpha}'_{-1,j} = \begin{cases} 1 & j = m_1 \\ 0 & \text{Otherwise} \end{cases}$$

for $i = 1$

$$\boldsymbol{\alpha}_{1,j} = \frac{1}{\sqrt{n_2+1}} \times \begin{cases} -\sqrt{n_2} & \text{for } j = m_1+1 \\ 1 & \text{for } j = m_1+2 \\ 0 & \text{Otherwise} \end{cases} \qquad \boldsymbol{\alpha}'_{1,j} = \begin{cases} 1 & j = m_1+1 \\ 0 & \text{Otherwise} \end{cases}$$

for $i = 2, \dots, m_2$

$$\boldsymbol{\alpha}_{i,j} = \begin{cases} -1/\sqrt{2} & \text{for } j = i+m_1 \\ 1/\sqrt{2} & \text{for } j = i+m_1+1 \\ 0 & \text{Otherwise} \end{cases} \qquad \boldsymbol{\alpha}'_{i,j} = \begin{cases} -1/\sqrt{2} & \text{for } j = m_1+i-1 \\ 1/\sqrt{2} & \text{for } j = m_1+i \\ 0 & \text{Otherwise} \end{cases}$$

Considering $\boldsymbol{\alpha}_i$ and $\boldsymbol{\alpha}'_i$ defined as above, $G$ and $G'$ are

$$G_{i,j} = \begin{cases} 1 & \text{for } i = j = -m_1, \ldots, m_2 \\ -1/2 & \text{for } i = j - 1 = -m_1, \ldots, m_2 - 1, \ i \neq -2, -1, 1 \\ -1/2 & \text{for } i = j + 1 = -m_1 + 1, \ldots, m_2, \ i \neq -1, 1, 2 \\ -1/\sqrt{2(n_1 + 1)} & \text{for } (i,j) = (-2,-1), (-1,-2) \\ -1/\sqrt{2(n_2 + 1)} & \text{for } (i,j) = (1,2), (2,1) \\ -\sqrt{n_1 n_2}/\sqrt{(1+n_1)(1+n_2)} & \text{for } (i,j) = (-1,1), (1,-1) \\ 0 & \text{Otherwise} \end{cases}$$

$$G'_{i,j} = \begin{cases} 1 & \text{for } i = j = -m_1, \ldots, m_2 \\ -1/2 & \text{for } i = j - 1 = -m_1, \ldots, m_2 - 1, \ i \neq -2, -1, 1 \\ -1/2 & \text{for } i = j + 1 = -m_1 + 1, \ldots, m_2, \ i \neq -1, 1, 2 \\ 1/\sqrt{2} & \text{for } (i,j) = (-2,-1), (-1,-2) \\ -1/\sqrt{2} & \text{for } (i,j) = (1,2), (2,1) \\ 0 & \text{Otherwise} \end{cases}$$